\documentclass[%
 reprint, amsmath, amssymb, aps
]{revtex4-1}

\usepackage{graphicx}
\usepackage{multirow}
\usepackage{dcolumn}
\usepackage{bm}

\begin{document}


\title{Single-shot 3D shape reconstruction using deep convolutional neural networks}%

\author{Hieu Nguyen}
\email{29nguyen@cua.edu}
\author{Zhaoyang Wang}%
\email{wangz@cua.edu}
\affiliation{Department of Mechanical Engineering, The Catholic University of America, Washington, D.C. 20064, USA}

\author{Hui Li}
\author{Qiang Qiu}
\affiliation{RvBust Ltd., 16 Keyuan Rd, Shenzhen, 518057, China}

\author{Yuzeng Wang}
\affiliation{School of Mechanical Engineering, Jinan University, Jinan, 250022, China}

\date{\today}

\begin{abstract}
A robust single-shot 3D shape reconstruction technique integrating the fringe projection profilometry (FPP) technique with the deep convolutional neural networks (CNNs) is proposed in this letter. The input of the proposed technique is a single FPP image, and the training and validation data sets are prepared by using the conventional multi-frequency FPP technique. Unlike the conventional 3D shape reconstruction methods which involve complex algorithms and intensive computation, the proposed approach uses an end-to-end network architecture to directly carry out the transformation of a 2D images to its corresponding 3D shape. Experiments have been conducted to demonstrate the validity and robustness of the proposed technique. It is capable of satisfying various 3D shape reconstruction demands in scientific research and engineering applications.
\end{abstract}

\maketitle

\section{Introduction}
Non-contact 3D shape reconstruction using the structured-light technique is commonly used in a wide range of applications including machine vision, reverse engineering, quality assurance, 3D printing, entertainment, etc. The technique typically retrieves the depth or height information with an algorithm based on geometric triangulation, where the structured light helps facilitate the required image matching or decoding process. According to the number of images required for each 3D reconstruction, the structured-light techniques can be classified into two categories: multi-shot and single-shot \cite{Li:16, Li:05, Zuo:18, Zhu:16, Cai:16}. The multi-shot techniques are good at capturing high-resolution 3D images at a limited speed, and the single-shot techniques are capable of acquiring 3D images at a fast speed to deal with dynamic scenes. Consequently, the multi-shot techniques are widely used as an industrial metrology for accurate shape reconstructions, whereas the single-shot ones receive tremendous attentions in the fields of entertainment and robotics. As technologies evolve at an ever-increasing pace, applying the concept of deep machine learning to the highly demanded single-shot 3D shape reconstruction has become feasible. This is the motivation of this letter.

In the machine learning field, deep convolutional neural networks (CNNs) have found numerous applications in object detection, image segmentation, image classification, scene understanding, medical image analysis, and natural language processing, etc. The recent advances of using the deep CNNs for image segmentation intend to make the network architecture become an end-to-end learning process. For example, Long \textit{et al.} \cite{Long:15} restored the downsampled feature map to the original size of the input using backwards convolution. An impressive network architecture, named UNet and proposed by Ronneberger \textit{et al.} \cite{Ronneberger:15}, extended the decoding path from Long's framework to yield a precise output with a relatively small number of training images. Similarly, Badrinarayanan \textit{et al.} \cite{Badrinarayanan:17} used an idea of upsampling the lowest of the encoder output to improve the resolution of the output with less computational resources.

In the CNN-based 3D reconstruction and depth detection applications, Eigen \textit{et al.} \cite{Eigen:14} and Liu \textit{et al.} \cite{Liu:15} respectively proposed a scheme to conduct the depth estimation from a single view using the CNNs. In their work, they used a third-party training data set produced by Kinect RGB-D sensors, which has low accuracy and is insufficient for a good learning. Inspired by these two methods, Choy \textit{et al.} \cite{Choy:16} proposed a novel architecture which employs recurrent neural networks (RNNs) among the autoencoder CNNs for single- and multi-view 3D reconstructions. Over the past year, the utilization of CNNs framework for fringe pattern analysis has been explored, such as pattern denoising and phase distribution determinations. For intance, Feng \cite{Feng:19} integrates the deep CNNs with three high-frequency patterns for 3D reconstruction and reliable phase unwrapping. Jeught \cite{Jeught:19} proposed a neural network with a large simulation dataset in the training process to acquire the height information of an object from a single-shot fringe pattern. A number of investigations \cite{Yan:19, Hao:19} have shown promising results on using the DNN models to improve the estimation and detection of phase distributions.

\begin{figure*}[!htbp]
\centering
	\includegraphics[width=185.0mm]{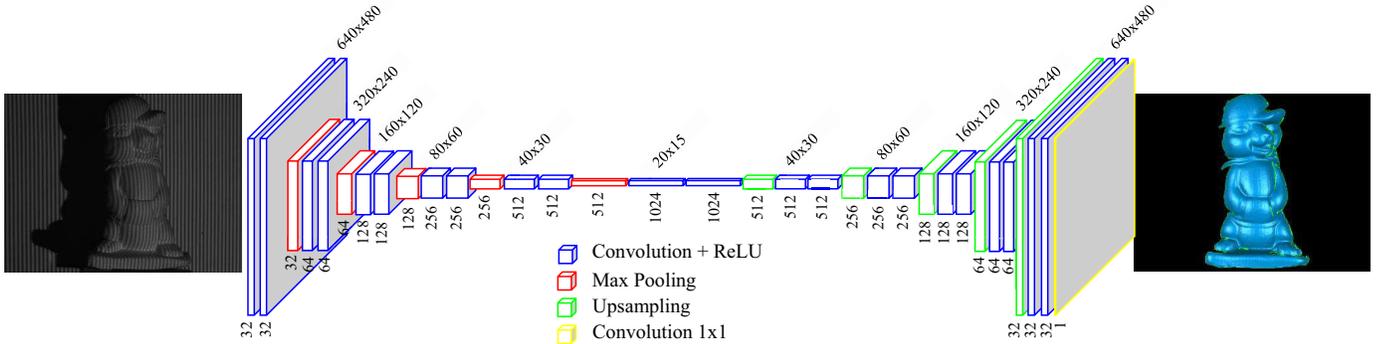}
\caption{Autoencoder architecture.}
\label{fig:fig1}
\end{figure*}

Based on the previous successful applications of the CNNs to image segmentation and 3D scene reconstruction, the exploration of utilizing the deep CNNs to accurately reconstruct the 3D shapes from a single structured-light image should be quite viable. With numerous parameters, a deep CNN model can be trained to approximate a very complex non-linear regressor that is capable of mapping a conventional structured-light image to its corresponding 3D depth or height map. At present, the robustness and importance of integrating the deep CNNs with one of the most widely used structured-light methods, fringe projection profilometry (FPP) technique, have not been fully recognized. In this letter, such an integration for accurate 3D shape reconstruction is investigated. The main idea is to transform a single-shot image, which has a high-frequency fringe pattern projected on the target, into a 3D image using a deep CNN that has a contracting encoder path and an expansive decoder path. Fig \ref{fig:fig1} demonstrates an example of the autoencoder-shaped model use in the proposed approach. Compared with the conventional 3D shape measurement techniques, the proposed technique is considerably simpler without using any geometric information or any complicated stereo-vision or triangulation-based computation. 

Using real and accurate training data is essential for a reliable machine learning model. Because the FPP technique is one of the most accurate 3D shape measurement techniques and is capable of performing 3D imaging with accuracy better than 0.1 mm, it is employed in this work to generate the required training and validation data for learning and the test data for evaluation. The proposed approach is elaborated below, starting from the training and validation data generation, and followed by the description of three deep CNNs.

\section{Methodology}
\subsection{Fringe projection profilometry technique for training data generation}
The most reliable FPP technique involves projecting a set of phase-shifted sinusoidal fringe patterns from a projector onto the objects, where the surface depth or height information is naturally encoded into the camera-captured fringe patterns for subsequent 3D reconstruction process. Technically, the fringe patterns help establish the correspondences between the captured image and the original reference image projected by the projector. In practice, the FPP technique determines the height or depth map from the phase distributions of the captured fringe patterns. The phase extraction process normally uses phase-shifted fringe patterns to calculate the fringe phase.

In general, the original reference fringes are straight, evenly spaced, and vertically (or horizontally) oriented. They are generated in a numerical way with the following function:
\begin{equation}
\label{fringe_generated}
 I_j (u,v) = I_0 \left[ 1+cos(\phi+\delta_j) \right ] = I_0 \left[ 1+cos(2\pi f\frac{u}{w}+\delta_j) \right ]
\end{equation}
where $I$ is the intensity value of the pattern at pixel coordinate $(u,v)$; the subscript $j$ denotes the $j$th phase-shifted image with $j=\left\{1,2, ...,m\right\}$, and m is the number of the phase-shift steps (e.g., $m=4$); $I_0$ is a constant coefficient indicating the value of intensity modulation; $f$ is the number of fringes in the pattern; $w$ is the width of the digital image; $\delta$ is the phase-shift amount; and $\phi$ is the fringe phase.

The fringe phase can be calculated from a standard four-step phase-shifting algorithm; however, such a phase value is wrapped in a range of 0 to $2\pi$ and must be unwrapped to obtain the true phase. In order to cope with the phase-unwrapping difficulty encountered in the cases of complex shapes and geometric discontinuities, a scheme of using multi-frequency fringe patterns is often employed by the FPP. The unwrapped phase distributions can be accordingly calculated from: 
\begin{equation}
\label{unwrap}
\phi_i (u,v) = \phi_i^w (u,v)+ INT \left( \frac{\phi_{i-1}\frac{f_i}{f_{i-1}}-\phi_i^w}{2\pi} \right )2\pi
\end{equation}
In the equation, the subscript $i$ indicates the $i$th fringe-frequency pattern with $i = \left\{2,3, ...,n\right\}$, and $n$ is the number of fringe frequencies; the superscript $w$ denotes the wrapped phase; $INT$ represents the function of rounding to the nearest integer; $f$ is again the number of fringes in the projection pattern, and $f_n > f_{n-1} > ...>f_1=1$. A practical example is $n = 4$ with $f_4 = 100$, $f_3 = 20$, and $f_2 = 4$.

The essential task of the FPP technique is to retrieve the out-of-plane depth or height map from the afore-calculated phase distributions of the highest frequency fringes. The governing equation for a generalized setup where the system components can be arbitrarily positioned \cite{Vo:12} is:
\begin{align}
\begin{split}
	\label{governing_equation}
	z &=\frac{\mathbf{C}\mathbf{p}^\intercal}{\mathbf{D}\mathbf{p}^\intercal} \\
	\mathbf{C} &= \left\{1  \ \ \, c_1\ \ \, c_2\ \ \, c_3\ \ \, \cdots\ \ \, c_9\ \ \, c_{10}\ \ \, c_{11} \right\} \\
	\mathbf{D} &= \left\{d_0\ \ \, d_1\ \ \, d_2\ \ \, d_3\ \ \, \cdots\ \ \, c_9\ \ \, d_{10}\ \ \, d_{11} \right\} \\
	\mathbf{p} &= \left\{1\ \ \, \phi\ \ \, u\ \ \, u\phi\ \ \, v\ \ \, v\phi\ \ \, u^2\ \ \, u^2\phi\ \ \, v^2\ \ \, v^2\phi\ \ \, uv\ \ \, uv\phi \right\}
\end{split}
\end{align}
where $z$ is the out-of-reference-plane height or depth at the point corresponding to the pixel $(u,v)$ in the captured images; $\phi$ is the unwrapped phase of the highest-frequency fringe pattern at the same pixel; and coefficients $c_1-c_{11}$ and $d_0-d_{11}$ can be determined by a calibration process using a flexible calibration board or a few gage blocks. Details can be found in \cite{Vo:12}.

\begin{figure*}[!htb]
\centering
	\includegraphics[width=180.0mm]{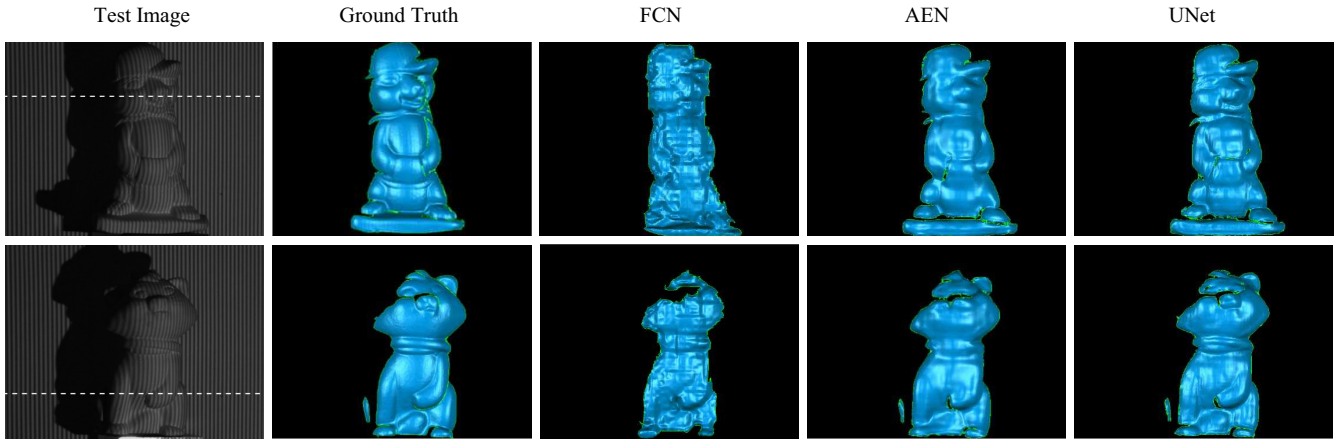}
\caption{3D reconstruction results of two representative test images.}
\label{fig:fig2}
\end{figure*}

As mentioned earlier, the FPP technique is employed not only to generate the training and validation data sets, but also to obtain the ground-truth results of the test data set for evaluation purpose. Once the training and validation data sets are available, they can be fed into a deep CNN model for the subsequent learning process. 

\subsection{Network architecture}

The adopted deep network is mainly made up of two components: the encoder path and the decoder path. Given a single fringe image of an object or objects, the network first encodes the input image into low-resolution feature maps. Then the network decodes the feature maps while upsampling them back to the original resolution with the final output as a 3D depth or height map. In the proposed approach, three deep CNNs are adopted, and they are described as follows:

\begin{itemize}
	\item \textbf{Fully convolutional networks} Fully convolutional network (FCN) is a well-known network that has been successfully applied to semantic segmentation. FCN adopts the encoder path from the contemporary classification networks (such as AlexNet, VGGNet, and GoogLeNet) and transforms the fully connected layers into convolution layers before upsampling the coarse output map to the same size as the input.
	\item \textbf{Autoencoder networks} The autoencoder network (AEN) has an encoder path and a symmetric decoder path. There are totally 33 layers, including 22 standard convolution layers, 5 max pooling layers, 5 transpose operation layers, and a $1 \times 1$ convolution layer. The AEN architecture is illustrated in Fig \ref{fig:fig1}.
	\item \textbf{UNet} The UNet is also a well-known network, and it has a similar architecture to the AEN. The key difference is that in the UNet the local context information from the encoder path are concatenated with the upsampled output, which can help increase the resolution of the final output.
\end{itemize}

In the learning process of the three CNNs, the training or validation data set is a four-dimensional array of size $s \times h \times w \times c$, where $s$ is the number of the training or validation samples; $h$ and $w$ are the spatial dimensions or the image dimensions; $c$ is the channel dimension, with $c=1$ for grayscale images and $c=3$ for color images. The networks contain convolution layers, pooling layers, transpose convolution layers, and unpooling layers; they do not contain any fully connected layers. Each convolution layer learns the local features from the input and produces the output features where the spatial axes of the output map remain the same but the depth axis changes following the convolution operation filters. A nonlinear activation function named rectified linear unit (ReLU), expressed as $max(0,x)$, is employed in each convolution layer. The max pooling layers with a $2 \times 2$ window and a stride of 2 are applied to downsample the feature maps through extracting only the max value in each window. In the AEN and UNet, the 2D transpose convolution layers are applied in the decoder path to transform the lower feature input back to higher resolution. Finally, a $1 \times 1$ convolution layer with one filter is attached to the final layer to transform the feature maps to the desired depth or height map. Unlike the conventional 3D shape reconstruction schemes that often require complex algorithms based on a profound understanding of techniques, the proposed approach depends on the numerous parameters in the networks, which are automatically trained, to play a vital role in the single-shot 3D reconstruction.

\section{Experiments and Results}
An experiment has been conducted to validate and demonstrate the capability of the proposed approach. The experiment uses a desktop computer with an  Intel Core i3-8100 3.6GHz processor, 16 GB RAM, and a Nvidia GeForce GTX 1070 graphics card. Keras, a popular Python deep learning library, is utilized to train the different CNN models. Nvidia's cuDNN deep neural network library is adopted to speed up the training process. The field of view of the experiment is about 155 mm, and the distance from the camera to the scene of interest is around 1.2 m. A number of small plaster sculptures serve as the objects, whose sizes and surface natures can help get reliable and high-accuracy 3D data sets.

\subsection{Training and test data acquisition}
As described previously, the multi-frequency FPP technique is employed to prepare the training, validation, and test data. The experiment uses four fringe frequencies (1, 4, 20, and 100) and the four-step phase-shifting schemes, which usually yield a good balance among accuracy, reliability, and capability. The first image of the last frequency (i.e., $f_4=100$) is chosen as the input image, and all other images are captured solely for the purpose of generating the ground-truth 3D height labels. Totally, there are 540, 60, and 72 samples in the training, validation, and test data sets, respectively; and each sample contains a single fringe image of object(s) and a corresponding height map. The data split ratio is roughly 80\%-10\%-10\% and is appropriate for such a case of small data sets. The data sets have been made publicly available for download \cite{Nguyen:19}. Visualization 1 shows the training and test data, where the left side displays the input images and the right side illustrates the corresponding ground truth. The background helps cover the field of view, but it is neither mandatory nor has to be flat. It is noted that the background are hidden in the visualization for better demonstration purpose, and the original shadow areas are excluded from the learning process.

\subsection{Training, analysis, and evaluation}
The training and validation data are applied to the learning process of the aforementioned three CNN models: FCN, AEN, and UNet. The optimization adopts a total of 300 epochs with a mini-batch size of 2 images. The learning rate is reduce by half whenever the validation loss does not improve within 20 consecutive epochs. A number of advanced regularization approaches has been implemented to tackle the over-fitting problem as well as to yield better performance, such as data augmentation, weight regularization, and dropout. Especially, the idea of using Grid Search method has been conducted to obtain the best hyperparameters for the training model. In order to check the network performance as the training process iterates, one of the test samples is randomly selected in advance. At the end of each epoch, a callback is performed on the selected test image to predict and save the intermediate result using the newly updated learning model (see Visualization 2). The learning can adopt either the binary cross-entropy or the mean squared error as the loss function. If the binary cross-entropy is selected, the ground truth data is normalized to a range between 0 and 1. 

The evaluation is carried out by calculating the errors of the reconstructed 3D shapes. Two errors, the mean relative error (MRE) and the root mean squared error (RMSE), are used in the analysis. Table 1 shows the performance errors of the three CNN models for single-shot 3D shape reconstruction. It can be seen that the FCN model yields the largest error among the three CNNs, and its learning time is also the longest because of the involved element-wise summation. The AEN model requires the least learning time, but its performance is slightly inferior to that of the UNet in terms of accuracy. 

\begin{figure}[!htb]
\centering
	\includegraphics[width=90.0mm]{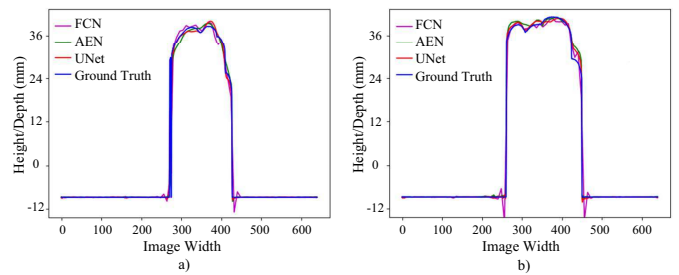}
\caption{Height distributions along a line in the representative test images.}
\label{fig:fig3}
\end{figure}

\begin{table}[!htb]
\caption{Performance evaluation of the CNNs.}
\centering
\begin{tabular}{|c|c|c|c|c|}
\hline
\multicolumn{2}{|c|}{Model}         & FCN      & AEN      & UNET     \\ \hline
\multicolumn{2}{|c|}{Training time} & 7 hrs    & 5 hrs    & 6 hrs    \\ \hline														
\multirow{2}{*}{Training}    & MRE  & 1.28e-3 & 8.10e-4 & 7.01e-4 \\ \cline{2-5} 
                             & RMSE (mm) & 1.47 & 0.80 & 0.71 \\ \hline
\multirow{2}{*}{Validation}  & MRE  & 1.78e-3 & 1.65e-3 & 1.47e-3 \\ \cline{2-5} 
                             & RMSE (mm) & 1.73 & 1.43 & 1.27 \\ \hline
\multirow{2}{*}{Test}        & MRE  & 2.49e-3 & 2.32e-3 & 2.08e-3 \\ \cline{2-5} 
                             & RMSE (mm)& 2.03 & 1.85 & 1.62 \\ \hline
\end{tabular}
\end{table}

\begin{figure*}[!htb]
\centering
	\includegraphics[width=180.0mm]{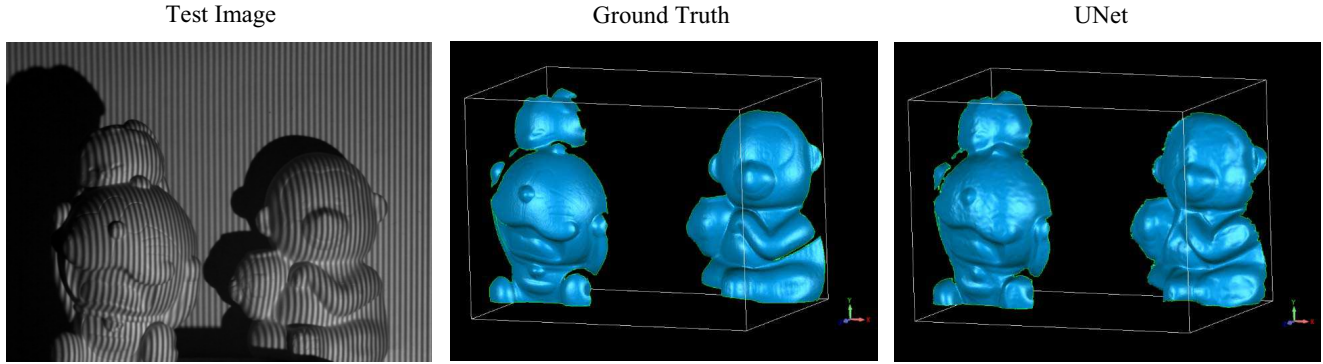}
\caption{3D reconstruction results of multiple separated objects.}
\label{fig:fig4}
\end{figure*}

Figure \ref{fig:fig2} demonstrates a visual comparison of the ground-truth 3D data and the reconstructed 3D data acquired with the three networks. The first image in each row is a representative input which captured an object with projected fringe patterns, and following the input image is the corresponding 3D ground-truth image. The next three images in each row are the reconstructed results from the deep learning with FCN, AEN, and UNet, respectively. In the reconstructed 3D figures, the background has been removed for better visualization purpose. Figure \ref{fig:fig3} shows the height distributions relative to the background plane along an arbitrary line highlighted in each of the initial input image. Again, it is evident from the Figs. \ref{fig:fig2} and \ref{fig:fig3} that the AEN and UNet models perform better than the FCN model. The main reason is that the FCN abruptly restores the high resolution feature map from the lower one, therefore, many details are lacking in the final reconstructed 3D results. The AEN and UNet each consists of its decoder path that is symmetric to the encoder path, which helps steadily propagate the context information between layers to produce features depicting detailed information. Unlike the AEN, the UNet contains concatenation operation to send extra local features to the decoder path. This handling helps the network to perform the best among the three networks, as can be seen from the figures.

Based on the successful reconstruction of an isolated object from the first dataset, a second dataset \cite{Nguyen:19} with multiple complex objects is captured and fed into the CNNs framework to check whether the proposed technique can solve the phase ambiguity, discontinuous surfaces issues as well as determine 3D depth information same as traditional technique. In the second dataset, there are 630, 70, and 40 samples in the training, validation, and test data sets, respectively. The UNet model with the best hyperparamters has been chosen for the training process of the second dataset. Figure \ref{fig:fig4} displays the multiple objects test image, the ground-truth depth reconstruction from FPP method, and the representative 3D reconstruction from UNet model. By visually comparing the result from the ground-truth shape and the predicted output shape, it can be seen that the Unet model successfully produces the accuracy output shape and solve the problems of phase ambiguity from multiple discontinuous objects even using only a single fringe image.

It is noteworthy that the 3D reconstruction time for a new single image is less than 50 ms on the aforementioned computer, which indicates that a real-time 3D shape reconstruction is practicable. Technically, the performance can be further improved with much larger training data sets as well as deeper networks. In practice, however, preparing a considerably large number of high-accuracy ground truth data is very time-consuming and challenging; furthermore, a deeper network will require a large amount of computer memory and computational time for the learning process. The future work can include improving the network model and preparing a larger data set as well as using less memory-consuming algorithms. 

\section{Conclusion}
In summary, a novel single-shot 3D shape reconstruction approach is presented. The approach employs three deep CNN models, including FCN, AEN and UNet, to quickly reconstruct the 3D shapes from a single image of the target with projected fringe patterns. The learning process is carried out through using the training and validation data acquired by a high-accuracy FPP technique. Experiments show that the UNet performs the best among the three networks. The validity of the approach gives great promise in the future research and development, which will include, but not limited to, using much larger data sets and large numbers of various objects as well as conducting a rigorous in-depth investigation on the CNN models.   

\section{Acknowledgment}
The authors thank Dr. Thanh Nguyen at The Catholic University of America for helpful discussion on the CNN models.

\bibliography{singleshotcnn}

\end{document}